%% file: thermoforce.tex
\documentclass[preprint,12pt]{elsarticle}

\usepackage{amssymb}
\usepackage{amsmath}
\usepackage{graphicx}
\usepackage{booktabs}
\usepackage{multirow}
\usepackage{xcolor}
\usepackage{rotating}
\usepackage[hidelinks]{hyperref}

\newcommand{\free}[1]{\textcolor{ForestGreen}{#1}}
\newcommand{\forced}[1]{\textcolor{BrickRed}{#1}}
\definecolor{ForestGreen}{RGB}{34,139,34}
\definecolor{BrickRed}{RGB}{178,34,34}

\newcommand{\placeholder}[2]{%
  \fbox{\parbox[c][#1][c]{0.92\linewidth}{\centering\itshape #2}}}

\journal{Energy and AI}

\begin{document}

\begin{frontmatter}

\title{ThermoForce: A Physics-Structured Interventional World Model
for Building HVAC Control}

\author{Yifan Wang\corref{cor1}}
\ead{yifan.wang18@mail.mcgill.ca}
\cortext[cor1]{Corresponding author.}

\affiliation{organization={Department of Mechanical Engineering, McGill University},
            addressline={817 Sherbrooke Street West},
            city={Montreal},
            postcode={H3A 2T7},
            state={QC},
            country={Canada}}

\begin{abstract}
Model predictive control (MPC) of building heating, ventilation, and
air-conditioning (HVAC) systems depends on a thermal model that can answer a
fundamentally causal question: what will the indoor temperature, energy use, and
comfort be \emph{if} a given control action is applied? Time-series foundation
models (TSFMs) now forecast passive building thermal trajectories with remarkable
zero-shot skill, and it is tempting to treat them as ready-made thermal models for
control. We show that this is unsafe: high factual forecasting accuracy does not
imply valid response to control interventions. An observational grey-box model
with the \emph{best} passive accuracy predicts the effect of cooling actions with
the wrong sign, and feeding control and weather covariates to a TSFM degrades
rather than improves its intervention response. We introduce ThermoForce, a
control-ready interventional thermal world model that keeps a TSFM frozen as a
passive free-response prior and learns a compact, physics-structured
forced-response operator for the causal effect of HVAC actuation. The operator is
monotone in the control input by construction, is identified from one to three
days of control excitation, and composes with the free response into a
counterfactual-capable world model. Across paired EnergyPlus heating and cooling
interventions, ThermoForce attains the lowest intervention-effect error and
correct effect sign where covariate-TSFM, observational grey-box, and
distillation baselines fail. Embedded in MPC on the BOPTEST benchmark, it reduces
thermal discomfort by 33--84\% relative to the native controller across three
two-week windows while simultaneously reducing energy, using a frozen backbone,
195 trainable parameters, and central-processing-unit-only computation.
ThermoForce reframes foundation models for building control: passive prediction
and forced intervention response must be structurally separated for a model to be
control-ready.
\end{abstract}

\begin{highlights}
\item Factual forecasting accuracy does not imply valid HVAC control response.
\item ThermoForce separates a frozen free-response prior from a forced-response operator.
\item The forced operator is monotone in the control input by construction.
\item It recovers intervention effects from one to three days of control excitation.
\item MPC deployment cuts BOPTEST discomfort 33--84\% while reducing energy.
\end{highlights}


\end{frontmatter}

\section{Introduction}
\label{sec:intro}

Buildings account for a large share of global final energy use, and their heating,
ventilation, and air-conditioning (HVAC) systems are the dominant controllable
load. Model predictive control (MPC) is the most successful advanced strategy for
operating these systems, repeatedly demonstrating energy and comfort improvements
over rule-based control~\cite{afram2014review,drgona2020allyouneed}. The
effectiveness of MPC hinges almost entirely on one component: a thermal model that
can predict how the building will respond to candidate control actions over a
planning horizon. The controller does not need to know only what the temperature
\emph{will be}; it needs to know what the temperature, energy, and comfort
\emph{would be} under each action it is considering. This is a causal,
counterfactual query, and it is what separates a control-ready model from an
ordinary forecaster.

Time-series foundation models (TSFMs) such as Chronos~\cite{ansari2024chronos} and
TimesFM~\cite{das2024timesfm} have recently transformed general-purpose
forecasting. Pretrained on enormous corpora of time series, they produce accurate
probabilistic forecasts zero-shot, and early studies confirm that they forecast
building temperature and load competitively with little or no target-building
data~\cite{park2025probabilistic}. It is therefore natural to ask whether a TSFM
can serve directly as the thermal model inside an HVAC controller. Doing so would
inherit the broad prior knowledge of the foundation model and sidestep the tedious
per-building identification that grey-box and physics-based models require.

We argue, and demonstrate, that this direct route is unsafe. A TSFM is a
\emph{passive} forecaster: it predicts how a system will evolve under the
statistical regularities of its history. HVAC control, however, is an
\emph{intervention}, not a passively observed covariate. Whether the model is
asked to extrapolate to a control action rarely seen in the target building's short
operating record, or to disentangle the control signal from the weather and
occupancy that co-vary with it, passive accuracy provides no guarantee that the
predicted \emph{effect of an action} is even directionally correct. Recent evidence
already hints at this: adding covariates to a TSFM does not reliably improve its
building thermal modeling~\cite{mulayim2025cantsfm}. The consequence for control has
not been made explicit, and it is severe. In our experiments an observational
grey-box model that achieves the best passive accuracy in a heating climate
predicts the effect of cooling actions with essentially inverted sign; a TSFM given
control and weather covariates forecasts intervention effects worse than the raw
TSFM. A controller built on such a model will confidently choose the wrong action.

The central message of this paper is therefore a reframing:
\begin{quote}
\emph{Factual forecasting accuracy is not sufficient for HVAC control. A thermal
model is control-ready only when it predicts the causal effect of control actions,
and this requires structural separation of passive evolution from forced
intervention response.}
\end{quote}

We operationalize this principle with \textbf{ThermoForce}, a control-ready
interventional thermal world model. ThermoForce decomposes the future zone
temperature into a \free{free response} and a \forced{forced response},
\begin{equation}
\underbrace{\hat{T}_{t+h}}_{\text{total response}}
= \underbrace{\free{\hat{T}^{\,\mathrm{free}}_{t+h}}}_{\text{frozen TSFM}}
+ \underbrace{\forced{\hat{T}^{\,\mathrm{forced}}_{t+h}}}_{\text{physics operator}},
\label{eq:superposition-intro}
\end{equation}
where the \free{free response} is the control-independent evolution driven by
weather, solar gains, occupancy, and thermal inertia, predicted by a
\emph{frozen} TSFM, and the \forced{forced response} is the additional effect of
HVAC actuation, predicted by a compact physics-structured operator. The
foundation model is left to do what it is genuinely good at, passive prediction,
and is never fine-tuned on the small, control-confounded target record. The
forced-response operator, by contrast, is small, identifiable, and constrained to
be monotone in the control input by construction, so the sign and boundedness of
control authority hold regardless of the learned weights. Their superposition is a
world model that answers counterfactual queries and drops directly into MPC
(Figure~\ref{fig:framework}).

Our contributions are as follows.
\begin{enumerate}
\item \textbf{The interventional-validity gap, quantified.} We show across paired
heating and cooling interventions that passive forecasting accuracy does not imply
control validity: the observational model with the best factual accuracy predicts
cooling effects with the wrong sign, and covariate-TSFM and distillation
surrogates remain unreliable. This reframes the modeling objective for building
control (Section~\ref{sec:results-interv}).
\item \textbf{A control-ready world model by free/forced superposition.} We keep a
TSFM frozen as a passive prior and add an explicit forced-response operator, so
that only the composition answers ``what if I apply this action'' queries
(Section~\ref{sec:method}).
\item \textbf{A monotone-by-construction interventional operator} with a stable
pole, sign-definite gain, and bounded state modulation, together with a two-stage
identification procedure that defeats thermostat-feedback confounding and recovers
control authority from one to three days of excitation (Sections~\ref{sec:operator}
and~\ref{sec:identification}).
\item \textbf{Closed-loop validation and efficiency.} Embedded in MPC on BOPTEST,
ThermoForce reduces thermal discomfort by 33--84\% relative to the native
controller across three two-week windows while reducing energy, using a frozen
backbone, 195 trainable parameters, and no graphics-processing-unit
(Section~\ref{sec:results-mpc}).
\end{enumerate}

\section{Related work}
\label{sec:related}

\paragraph{Time-series foundation models.}
TSFMs pretrain a single sequence model on large heterogeneous time-series corpora
and forecast new series zero-shot. Chronos tokenizes scalar values and trains a
language-model architecture for probabilistic
forecasting~\cite{ansari2024chronos}; TimesFM is a decoder-only model pretrained on
billions of time points~\cite{das2024timesfm}. These models are powerful passive
priors, but they are trained and evaluated as forecasters, not as models of
response to interventions.

\paragraph{Foundation models for building energy.}
TSFMs have been examined for building energy systems, mostly for forecasting and
representation. Probabilistic forecasting and fine-tuning of TSFMs for building
energy systems has shown competitive accuracy~\cite{park2025probabilistic}, while a
systematic study reports limited zero-shot generalizability and, critically, that
adding covariates does not necessarily improve thermal behavior
modeling~\cite{mulayim2025cantsfm}. Our work builds on this observation and shows
its decisive consequence for control: the missing capability is not another
forecasting benchmark but validity under intervention.

\paragraph{Physics-constrained thermal modeling and MPC.}
Grey-box resistance-capacitance (RC) models and physics-constrained neural networks
are standard building thermal models for MPC. Physics-constrained deep learning
improves the physical consistency of multi-zone thermal
models~\cite{drgona2021physics}, and physically consistent deep learning has been
combined with MPC for energy optimization~\cite{xiao2023physically}. Economic MPC
studies compare model-based and data-driven approaches on standardized
benchmarks~\cite{zheng2024economic}. ThermoForce is a hybrid of a frozen learned
prior and an explicit physical forced-response structure, and it targets
intervention validity rather than only passive fit.

\paragraph{Differentiable and learned control.}
Differentiable predictive control (DPC) learns an explicit neural control law by
backpropagating an economic MPC objective through a differentiable closed-loop
model~\cite{drgona2022dpc}. DPC requires a differentiable plant model; ThermoForce
provides exactly such an identifiable and intervention-valid world model. We
therefore position ThermoForce as complementary to DPC rather than as a competing
controller.

\paragraph{Distillation of foundation models.}
Closest to our setting, ThermoStill distills a TSFM into a compact thermal-dynamics
model for HVAC MPC~\cite{liang2026thermostill}. ThermoForce takes the opposite
structural stance: rather than compressing the TSFM into one dynamics network, it
keeps the TSFM frozen for the free response and isolates the control effect in an
explicit forced-response operator. Table~\ref{tab:positioning} positions
ThermoForce against these lines of work, and Table~\ref{tab:thermostill} contrasts
it with ThermoStill directly. As we show empirically, a distilled single-network
surrogate is unstable at recovering control effects from sparse excitation, whereas
the explicit decomposition is stable and sign-correct.

\begin{table}[t]
\centering
\caption{Positioning of ThermoForce against related lines of work. A check mark
indicates the property is present.}
\label{tab:positioning}
\small
\input{tables/tab_positioning}
\end{table}

\begin{table}[t]
\centering
\caption{Direct contrast with ThermoStill~\cite{liang2026thermostill}, the closest
foundation-model approach to HVAC MPC.}
\label{tab:thermostill}
\small
\input{tables/tab_thermostill}
\end{table}

\section{Problem formulation}
\label{sec:problem}

Let $T_t$ denote the zone air temperature at time $t$, $u_t \in [0,1]$ the
normalized HVAC control input (heating or cooling actuation), and
$w_t=(T^{\mathrm{out}}_t, I^{\mathrm{sol}}_t, \dots)$ the exogenous passive drivers
(outdoor temperature, solar irradiance, and time of day). Given a history
$\mathcal{H}_t=\{T_\tau,u_\tau,w_\tau\}_{\tau\le t}$, a control-oriented thermal
model must produce, for any \emph{planned} control trajectory $u_{t:t+H}$ over a
horizon $H$, a forecast of the resulting temperature trajectory,
\begin{equation}
\hat{T}_{t+h}(u_{t:t+h}) = \Phi\!\left(\mathcal{H}_t,\, w_{t:t+h},\, u_{t:t+h}\right),
\qquad h=1,\dots,H.
\label{eq:model}
\end{equation}

We distinguish two queries. The \emph{factual} query evaluates
Eq.~\eqref{eq:model} at the control that was actually applied,
$\hat{T}^{\mathrm{fact}}_{t+h}=\hat{T}_{t+h}(u^{\mathrm{obs}}_{t:t+h})$, and is
scored by standard forecast error. The \emph{counterfactual} (interventional) query
evaluates Eq.~\eqref{eq:model} at an alternative control $u'_{t:t+h}$, and the
quantity of interest is the \emph{intervention effect}
\begin{equation}
\widehat{\Delta}_{t+h}(u',u) = \hat{T}_{t+h}(u'_{t:t+h}) - \hat{T}_{t+h}(u_{t:t+h}).
\label{eq:effect}
\end{equation}
An MPC controller ranks candidate actions using precisely
Eq.~\eqref{eq:effect}; a model can be accurate on the factual query yet arbitrarily
wrong on the interventional one. The goal of this paper is a model whose
$\widehat{\Delta}$ is accurate and directionally correct, i.e., a
\emph{control-ready} model, from little target-building data.

\section{Method: ThermoForce}
\label{sec:method}

\begin{figure}[t]
\centering
\placeholder{7.0cm}{\textbf{Figure 1 (framework overview) --- to be drawn.}\\[4pt]
Conceptual schematic of ThermoForce. Left: the observed zone temperature is split
into a control-free component and an accumulated forced state. Center-top: a
\emph{frozen} time-series foundation model forecasts the free response (passive
prior); center-bottom: a compact physics-structured operator predicts the forced
response of HVAC actuation. Right: their superposition forms a counterfactual world
model that answers ``what if control $u$ is applied?'' and is queried by MPC over
candidate action trajectories. Suggested width: full text width; height
approximately 7\,cm.}
\caption{Overview of the ThermoForce interventional world model. The frozen
foundation model supplies the passive free response; a small monotone operator
supplies the forced response; their superposition is control-ready and drives MPC.}
\label{fig:framework}
\end{figure}

\subsection{Free/forced superposition}
\label{sec:superposition}

Consider a linear time-invariant approximation of the one-step zone dynamics within
a short window,
\begin{equation}
T_{t+1} = a\,T_t + g\,u_t + c\,T^{\mathrm{out}}_t + d\,I^{\mathrm{sol}}_t + e,
\label{eq:arx}
\end{equation}
with stable pole $a\in(0,1)$ and control gain $g$. Define the \forced{forced state}
$F$ by $F_{t_0}=0$ and $F_{t+1}=a F_t + g u_t$, i.e., the accumulated effect of the
control sequence alone. Subtracting, the \free{free component}
$T^{\mathrm{free}}_t = T_t - F_t$ evolves as
\begin{equation}
T^{\mathrm{free}}_{t+1} = a\,T^{\mathrm{free}}_t + c\,T^{\mathrm{out}}_t
+ d\,I^{\mathrm{sol}}_t + e,
\label{eq:free-evol}
\end{equation}
which carries \emph{no} control influence. This exact decomposition motivates the
architecture: a univariate forecaster applied to the control-free series
$T^{\mathrm{free}}$, plus a rollout of the forced state $F$ under a \emph{planned}
control $u$, yields a counterfactual-capable composite forecast. Concretely, given
context temperatures and applied controls we compute
$(T^{\mathrm{free}}_{\text{ctx}}, F_{\mathrm{end}})$ by
Eq.~\eqref{eq:free-evol}, forecast the free series with a frozen TSFM, and compose
\begin{equation}
\hat{T}_{t+h} = \free{\underbrace{f^{\mathrm{TSFM}}_{\theta}
\!\left(T^{\mathrm{free}}_{\text{ctx}}\right)_h}_{\text{free response}}}
\;+\;
\forced{\underbrace{F_h\!\left(u_{t:t+h}; F_{\mathrm{end}}\right)}_{\text{forced response}}}.
\label{eq:compose}
\end{equation}
The frozen TSFM $f^{\mathrm{TSFM}}_{\theta}$ has parameters $\theta$ that are never
updated. Only the forced-response operator is learned on the target building.

\subsection{Monotone interventional operator}
\label{sec:operator}

The forced response is produced by a compact operator that generalizes the linear
term $g u_t$ to a state-modulated gain while preserving physical structure. The
forced state evolves as
\begin{equation}
\forced{F_{t+1} = a\,F_t + g\,m(x_t)\,u_t},
\label{eq:operator}
\end{equation}
with
\begin{align}
a &= a_{\mathrm{lo}} + (a_{\mathrm{hi}}-a_{\mathrm{lo}})\,\sigma(\tilde a),
& &\text{(stable pole)}\label{eq:pole}\\
g &= \pm\,\mathrm{softplus}(\tilde g),
& &\text{(sign-definite gain)}\label{eq:gain}\\
m(x) &= 1 + \gamma\,\tanh\!\big(\mathrm{MLP}(x)\big),\quad \gamma<1,
& &\text{(bounded modulation)}\label{eq:mod}
\end{align}
where $\sigma$ is the logistic function, the sign of $g$ is positive for heating and
negative for cooling, and $x_t$ is a normalized feature vector of the zone and
weather state (excluding $u$). Because the modulation excludes the control input,
the marginal effect of the control on the forced state $k$ steps ahead is
\begin{equation}
\frac{\partial F_{t+k}}{\partial u_t} = a^{k}\,g\,m(x_t),
\label{eq:marginal}
\end{equation}
which is \emph{sign-definite} and bounded, since $m(x)\in(1-\gamma,1+\gamma)>0$ and
$a>0$. We state this as a proposition.

\begin{quote}
\textbf{Proposition (monotone control authority).} For any multilayer-perceptron
weights and any $\gamma<1$, the operator in Eq.~\eqref{eq:operator} is monotone in
the control input: $\partial F_{t+k}/\partial u_t$ has the fixed sign of $g$ and
magnitude bounded in $\big[a^{k} g (1-\gamma),\, a^{k} g (1+\gamma)\big]$.
\end{quote}

\noindent This is exactly the property that observational fits violate: a model may
fit the factual data well yet imply a control response of the wrong sign or
unbounded magnitude. ThermoForce guarantees correct, bounded control authority by
construction, which is what makes it safe to place inside a controller.

\subsection{Identification under feedback confounding}
\label{sec:identification}

Operational building data is closed-loop: the control $u$ is itself a function of
the temperature it regulates. A naive joint least-squares fit of
Eq.~\eqref{eq:arx} can therefore explain the data with $g\approx 0$, attributing all
dynamics to the free channel. We defeat this with a two-stage estimator. In the
first stage, HVAC-off samples ($u\le u_{\mathrm{off}}$) identify the free
parameters $(a,c,d,e)$ under physical box constraints, since the dynamics are
purely passive when the actuator is idle. In the second stage, with the free
parameters fixed, the HVAC-on residual is regressed on $u$ to estimate the
sign-clipped gain $g$. For counterfactual identification we additionally use
\emph{paired} probe episodes, fitting the operator on the difference between two
runs that share weather and occupancy but differ in control,
\begin{equation}
D_{t+1} = a\,D_t + g\,m(x^{A}_t)\,u^{A}_t - g\,m(x^{B}_t)\,u^{B}_t,
\label{eq:paired}
\end{equation}
where $D=T^{A}-T^{B}$ is the observed paired temperature difference. This isolates
the causal effect of the control change from all shared confounders and requires
only one to three probe days.

\subsection{Deployment in model predictive control}
\label{sec:deployment}

At each control step the composite model of Eq.~\eqref{eq:compose} is queried for a
set of candidate control trajectories over the planning horizon, and the trajectory
minimizing a weighted energy-plus-discomfort objective subject to comfort bounds is
selected; its first action is applied, and the horizon recedes. Because the free
response is supplied by the frozen TSFM and only the small forced operator is
building-specific, the entire controller runs without a graphics-processing-unit.

\section{Experimental protocol}
\label{sec:protocol}

\paragraph{Datasets and simulators.}
We evaluate on three complementary sources (Table~\ref{tab:datasets}). REFIT
provides real residential temperature records for few-shot factual forecasting.
EnergyPlus~\cite{crawley2001energyplus} generates paired counterfactual
interventions: for each cell we run bitwise-identical factual and perturbed
simulations that differ only in the setpoint schedule, so paired differences are
pure causal effects. BOPTEST~\cite{blum2021boptest} provides a high-fidelity
closed-loop benchmark with standardized key performance indicators for control.

\begin{table}[t]
\centering
\caption{Datasets, simulators, and their role in the evaluation.}
\label{tab:datasets}
\small
\begin{tabular}{p{2.6cm}p{2.2cm}p{2.0cm}p{4.2cm}}
\toprule
Source & Type & Scale & Role \\
\midrule
REFIT & Real homes & 19 houses, 30-min & Few-shot factual forecasting \\
EnergyPlus paired interventions & Simulation (setpoint counterfactuals) & 6 cells (2 climates $\times$ 3 windows), 15-min & Interventional validity (heating and cooling) \\
BOPTEST hydronic heat pump & Co-simulation benchmark & 3 two-week windows, 1-h control & Closed-loop MPC key performance indicators \\
\bottomrule
\end{tabular}
\end{table}

\paragraph{Intervention suite.}
The EnergyPlus suite spans setpoint shifts of $\pm 1,\pm 2,\pm 3\,^{\circ}$C over
three daily windows (morning, afternoon, evening) in a cold heating climate
(Chicago) and a hot cooling climate (Phoenix), giving six cells, each aggregating
44 held-out paired windows. Statistical tests are performed at the cell level
($n=6$) to avoid treating overlapping windows as independent samples.

\paragraph{Metrics.}
For factual forecasting we report root-mean-square error (RMSE) and per-building
win rate. For interventional validity we report the intervention-effect RMSE (the
error of $\widehat{\Delta}$ in Eq.~\eqref{eq:effect}) and the sign accuracy (the
fraction of steps at which the predicted effect has the correct sign). For
closed-loop control we report the BOPTEST energy, thermal discomfort (in
Kelvin-hours), and operating cost.

\paragraph{Validation-safe selection.}
All hyperparameters, guard thresholds, and method-selection rules are chosen on
validation or warmup data only; the test data is scored once. The modulation
coefficient is fixed a priori at $\gamma=0.75$.

\paragraph{Implementation.}
The frozen backbone is Chronos-Bolt-small (48\,M parameters, never updated). The
operator uses a two-layer multilayer perceptron (195 trainable parameters), trained
with Adam for 300--400 epochs. All computation is central-processing-unit-only.

\section{Results I: interventional validity}
\label{sec:results-interv}

We begin with the paper's central experiment, because it defines what
control-readiness means. Table~\ref{tab:interv} reports factual accuracy alongside
interventional validity for every method, separated by heating and cooling.

\begin{table}[t]
\centering
\caption{Interventional validity versus factual accuracy on the paired EnergyPlus
suite (mean over cells). Factual RMSE measures passive skill; effect RMSE and sign
accuracy measure control validity. Monotonicity violation was zero for all methods
on all cells and is omitted.}
\label{tab:interv}
\small
\input{tables/tab_intervention_master}
\end{table}

\begin{figure}[t]
\centering
\includegraphics[width=0.72\linewidth]{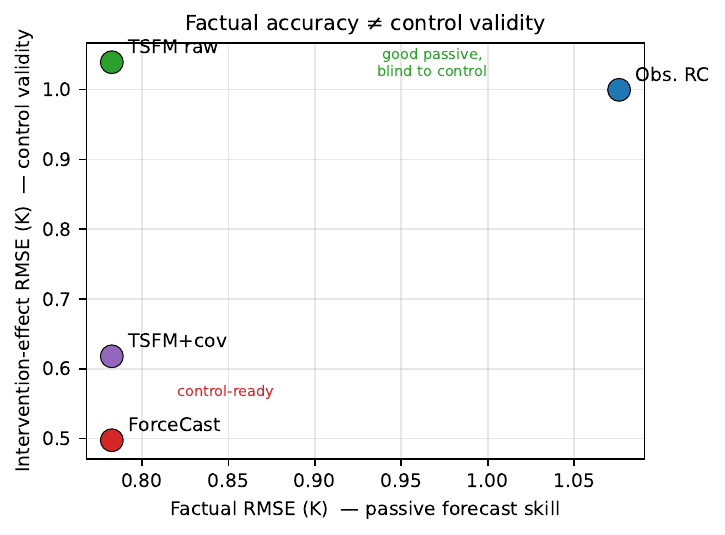}
\caption{Factual accuracy does not imply control validity. Each marker is a method
averaged over the intervention cells; the horizontal axis is passive forecast error
and the vertical axis is intervention-effect error. The observational grey-box model
(top right) is accurate passively yet worst on control validity; ThermoForce
(bottom) is strong on both.}
\label{fig:scatter}
\end{figure}

The result is decisive and, at first sight, paradoxical. In the heating climate the
observational grey-box model attains the \emph{best} factual RMSE (0.761\,K), better
than ThermoForce, yet its intervention-effect RMSE (1.360\,K) is nearly twice as
large as ThermoForce's (0.751\,K). In the cooling climate the same model is not
merely inaccurate on effects, it is \emph{directionally wrong}: its sign accuracy
collapses to 0.01, meaning it predicts that cooling actions warm the zone. Feeding
control and weather covariates to the TSFM improves its effect estimate over the raw
TSFM but remains well behind ThermoForce, and a distillation-style surrogate is
unstable, with per-cell effect RMSE ranging from 0.19 to 9.73\,K as few-shot
distilled rollouts diverge. ThermoForce achieves the lowest effect RMSE in both
climates (0.751\,K heating, 0.245\,K cooling) and high sign accuracy (0.85 and
0.95). Figure~\ref{fig:scatter} visualizes the decoupling: methods can be excellent
on the horizontal (passive) axis while failing on the vertical (control) axis, and
only ThermoForce is strong on both.

A one-sided Wilcoxon signed-rank test at the cell level confirms the ranking:
ThermoForce has lower effect RMSE than the raw TSFM and the observational model on
all six cells ($p=0.016$ each), and lower than the covariate TSFM on five of six
cells. The monotone construction earns its place in the cooling climate, where it
improves the effect RMSE from 0.331\,K (unconstrained gain) to 0.245\,K; in the
heating climate the two are statistically comparable, consistent with the
interpretation that monotonicity is a robustness guarantee that matters most under
sign-sensitive extrapolation.

\section{Results II: factual forecasting and covariate baselines}
\label{sec:results-factual}

Having established control validity, we confirm that ThermoForce does not sacrifice
passive forecasting skill. Table~\ref{tab:refit} reports few-shot factual RMSE on
REFIT with a three-day identification budget. ThermoForce improves the six-hour
median RMSE from 0.458\,K (raw TSFM) to 0.404\,K and wins on the majority of homes,
a statistically significant improvement (one-sided Wilcoxon, $p=0.0024$). At the
24-hour horizon ThermoForce is statistically comparable to the raw TSFM, preserving
the strong foundation-model prior without degrading it. Both are far ahead of the
grey-box RC baseline.

\begin{table}[t]
\centering
\caption{Few-shot factual forecasting on REFIT (median RMSE over 19 homes,
three-day budget). ThermoForce improves the six-hour horizon and preserves the
24-hour horizon.}
\label{tab:refit}
\small
\input{tables/tab_main_refit}
\end{table}

Crucially, the naive alternative of feeding covariates to the TSFM does not help.
Table~\ref{tab:cov} shows that a validation-tuned covariate TSFM, even when granted
oracle future weather, forecasts \emph{worse} than the zero-shot TSFM, and adding an
autoregressive term degrades it further; ThermoForce wins on 15 of 19 homes at six
hours. This corroborates the finding of~\cite{mulayim2025cantsfm} and reinforces the
central thesis: control-readiness requires structural separation of passive and
forced response, not merely a richer input to a monolithic forecaster.

\begin{table}[t]
\centering
\caption{Covariate-TSFM baselines versus ThermoForce on REFIT (median RMSE, K=3
days). Feeding covariates degrades the foundation-model forecast.}
\label{tab:cov}
\small
\input{tables/tab_covariate}
\end{table}

\begin{figure}[t]
\centering
\includegraphics[width=0.6\linewidth]{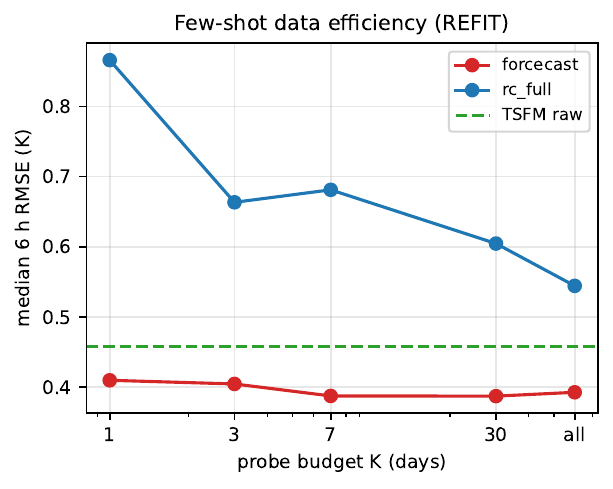}
\caption{Few-shot data efficiency on REFIT. ThermoForce is below the zero-shot TSFM
from a single identification day; the grey-box RC model requires far more data.}
\label{fig:budget}
\end{figure}

\begin{figure}[t]
\centering
\includegraphics[width=0.85\linewidth]{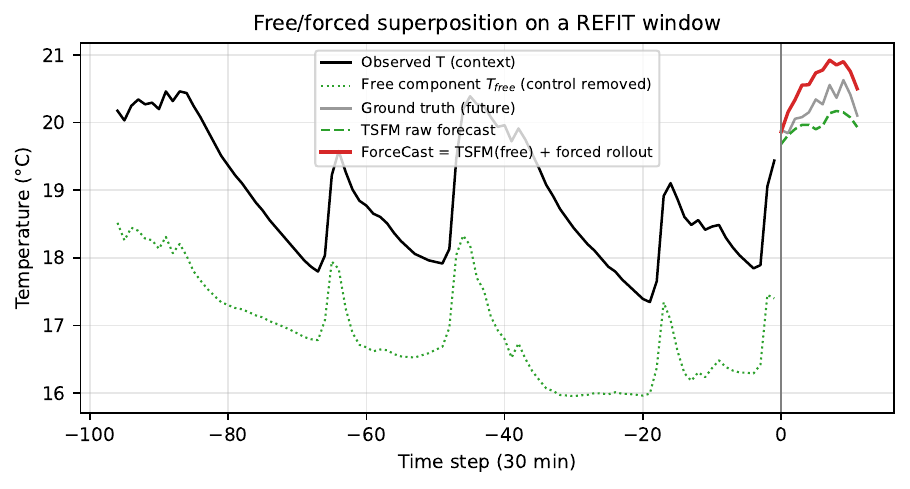}
\caption{Free/forced superposition on a REFIT window. The control-free component
(green) is smoother and lower than the observed temperature; the ThermoForce
composite forecast (red) tracks the ground truth more closely than the raw TSFM.}
\label{fig:decomp}
\end{figure}

\begin{figure}[t]
\centering
\includegraphics[width=0.85\linewidth]{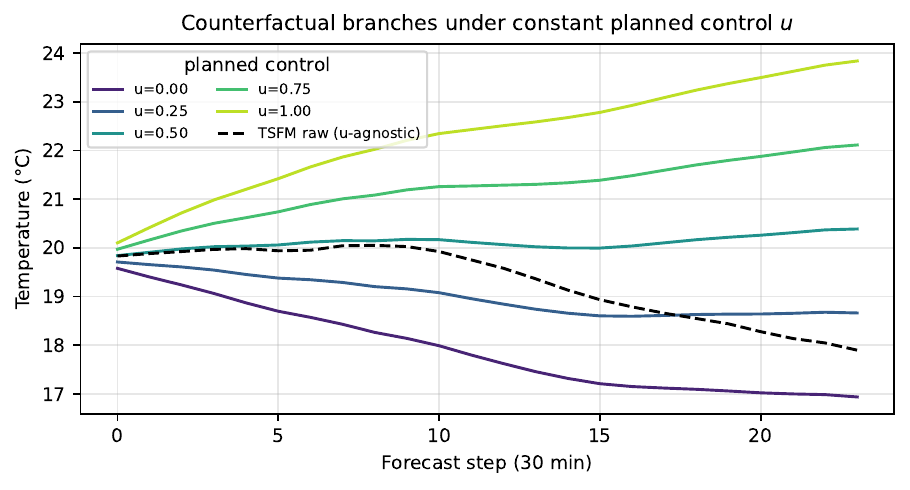}
\caption{Counterfactual capability. ThermoForce produces distinct temperature
trajectories for planned control levels $u\in\{0,0.25,0.5,0.75,1\}$ (colored),
whereas the raw TSFM (dashed) is insensitive to the control action.}
\label{fig:cf}
\end{figure}

Figure~\ref{fig:budget} shows the data efficiency: ThermoForce is below the raw
TSFM from a single identification day and remains flat thereafter, whereas the
grey-box model needs far more data and never catches up.
Figure~\ref{fig:decomp} illustrates the free/forced decomposition on a
representative window, and Figure~\ref{fig:cf} shows the counterfactual capability
directly: ThermoForce produces distinct temperature trajectories for different
planned control levels, while the raw TSFM is insensitive to the control input.

\section{Results III: closed-loop control}
\label{sec:results-mpc}

We deploy ThermoForce as the world model inside MPC on the BOPTEST hydronic
heat-pump benchmark and evaluate three full two-week windows against the native
BOPTEST controller and an RC-MPC controller that uses the same forced operator
without the TSFM free response. Table~\ref{tab:closedloop} reports the key
performance indicators.

\begin{table}[t]
\centering
\caption{Closed-loop key performance indicators on BOPTEST (three two-week
hydronic heat-pump windows). ThermoForce-MPC reduces discomfort and energy relative
to the native controller in every window.}
\label{tab:closedloop}
\small
\input{tables/tab_closedloop}
\end{table}

\begin{figure}[t]
\centering
\includegraphics[width=0.7\linewidth]{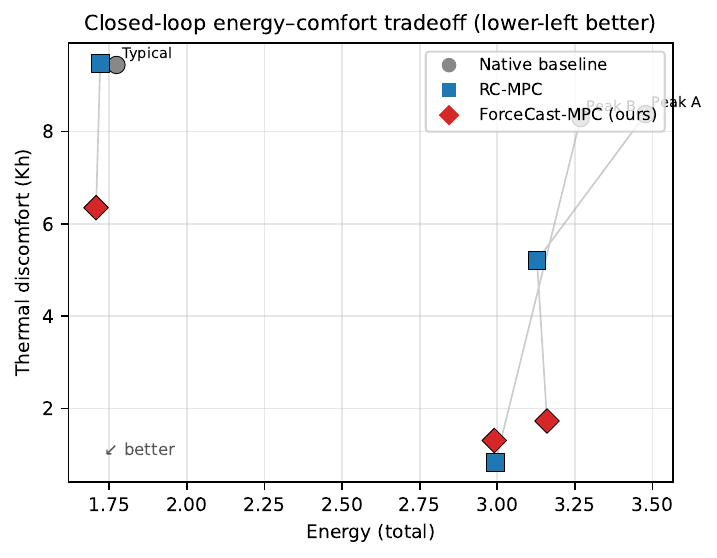}
\caption{Closed-loop energy-comfort tradeoff on BOPTEST. Each line connects the
native controller, RC-MPC, and ThermoForce-MPC for one two-week window; lower-left
is better. ThermoForce-MPC dominates the native controller in every window.}
\label{fig:pareto}
\end{figure}

\begin{figure}[t]
\centering
\includegraphics[width=0.95\linewidth]{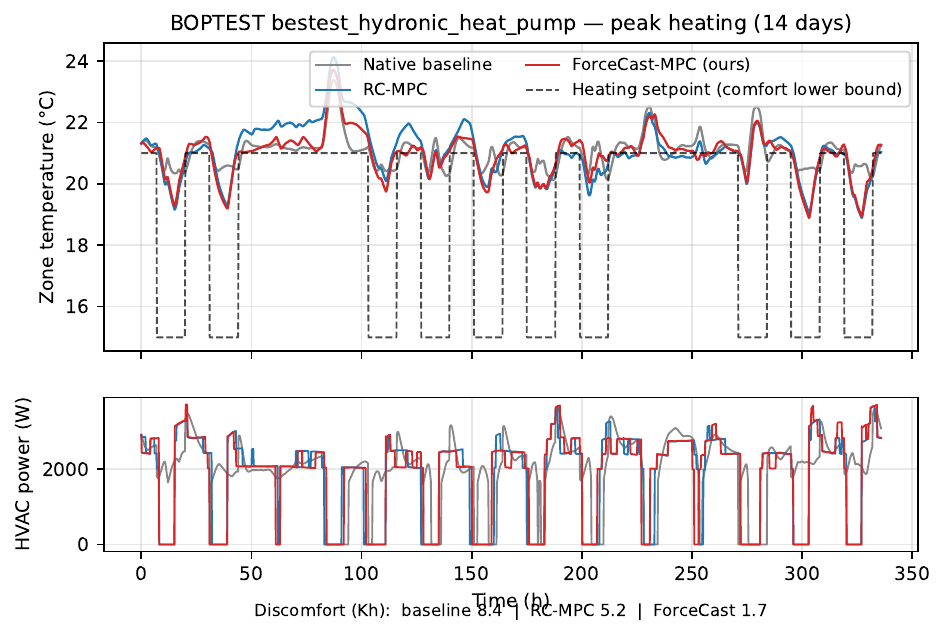}
\caption{Closed-loop operation on the BOPTEST hydronic heat-pump peak window. Top:
zone temperature versus the heating comfort bound. Bottom: HVAC power. ThermoForce-MPC
maintains comfort with lower accumulated discomfort than the native controller.}
\label{fig:cl}
\end{figure}

Across all three windows ThermoForce-MPC reduces thermal discomfort by 33--84\%
relative to the native controller while simultaneously reducing energy by 4--9\%. It
reduces discomfort far below the native baseline in every case and is competitive
with RC-MPC on energy. Figure~\ref{fig:pareto} plots the energy-comfort tradeoff:
ThermoForce-MPC sits in the favorable lower-left region across all windows,
dominating the native controller. Figure~\ref{fig:cl} shows the peak-window zone
temperature and heat-pump power, illustrating how ThermoForce-MPC tracks the comfort
lower bound more tightly than the native controller.

\section{Results IV: ablation and robustness}
\label{sec:ablation}

\paragraph{Operator ablation.}
Table~\ref{tab:ablation} isolates the contribution of the monotone-neural gain
against a linear gain on paired-difference effect prediction. The neural operator
matches the linear one at very small budgets and improves clearly at seven and
thirty probe days, confirming that the state modulation adds value once sufficient
excitation is available while never harming the few-shot regime.

\begin{table}[t]
\centering
\caption{Operator ablation: linear versus monotone-neural gain (paired-difference
effect RMSE, K). The neural gain improves with probe budget.}
\label{tab:ablation}
\small
\input{tables/tab_ablation_operator}
\end{table}

\begin{figure}[t]
\centering
\includegraphics[width=0.6\linewidth]{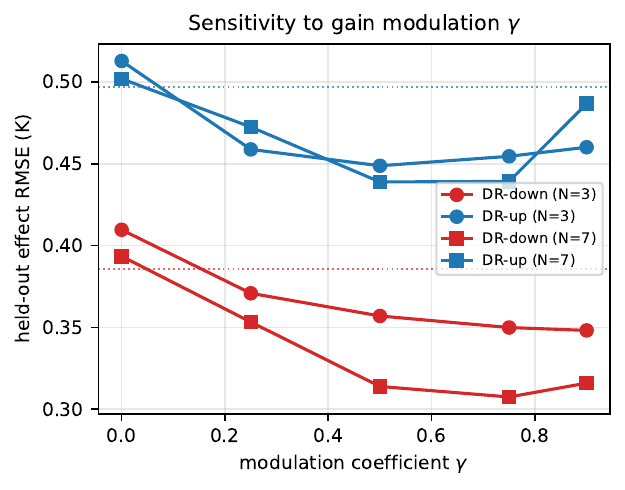}
\caption{Sensitivity to the modulation coefficient $\gamma$. The effect RMSE
improves from the linear case and plateaus, indicating a broad optimum around the
default $\gamma=0.75$.}
\label{fig:gamma}
\end{figure}

\paragraph{Hyperparameter sensitivity.}
Figure~\ref{fig:gamma} sweeps the modulation coefficient $\gamma$. Performance
improves from the linear case ($\gamma=0$) and plateaus over a wide range, so the
default $\gamma=0.75$ sits in a broad optimum; the method is not brittle to this
choice.

\begin{figure}[t]
\centering
\includegraphics[width=0.6\linewidth]{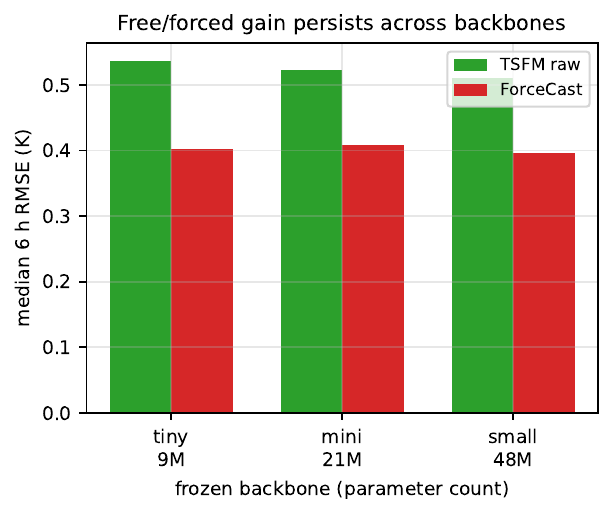}
\caption{The free/forced gain persists across frozen backbone scales
(tiny, mini, small).}
\label{fig:backbone}
\end{figure}

\paragraph{Backbone generality.}
Table~\ref{tab:backbone} swaps the frozen backbone across three Chronos scales
spanning a fivefold parameter range. ThermoForce improves over its own raw TSFM on
every home for every backbone, confirming that the free/forced gain is a property of
the decomposition rather than of one model. Figure~\ref{fig:backbone} visualizes
this consistency.

\begin{table}[t]
\centering
\caption{Backbone generality (median six-hour RMSE over eight REFIT homes, K=3).
The gain persists across backbone scales.}
\label{tab:backbone}
\small
\input{tables/tab_backbone}
\end{table}

\paragraph{Computational efficiency.}
Table~\ref{tab:efficiency} summarizes the cost. The interventional operator has 195
trainable parameters, per-building identification takes tens of milliseconds to a
few seconds, and inference is milliseconds per forecast, all on a
central-processing-unit. ThermoForce is thus control-ready with negligible online
cost and no backbone fine-tuning.

\begin{table}[t]
\centering
\caption{Computational cost (central-processing-unit only).}
\label{tab:efficiency}
\small
\input{tables/tab_efficiency}
\end{table}

\section{Discussion}
\label{sec:discussion}

\paragraph{Why not simply fine-tune the TSFM?}
Fine-tuning a foundation model on the target building can improve factual accuracy,
but it does not guarantee valid intervention response and risks overfitting the
sparse, feedback-confounded control excitation available in operation. ThermoForce
deliberately freezes the passive prior and learns only a small structured
forced-response operator, so its objective is control-valid counterfactuals under
limited data rather than in-distribution forecast accuracy. The covariate-TSFM
result (Table~\ref{tab:cov}) shows concretely that simply exposing the control
signal to a monolithic model is not enough.

\paragraph{Relation to learned controllers.}
ThermoForce is a world model, not a controller. It is complementary to
differentiable predictive control and reinforcement learning, which both require a
plant model: ThermoForce supplies an identifiable, interpretable, and
intervention-valid plant that such methods can use, and its counterfactual outputs
are directly inspectable for commissioning and safety review.

\paragraph{Practical implications.}
Because the method is central-processing-unit-only, uses a frozen backbone, and
identifies from one to three days of excitation, it is deployable on standard
building automation hardware and adapts quickly to a new building or season. The
monotone construction provides a certificate that control authority is
directionally correct, which is a prerequisite for trust in any model placed inside
a controller.

\section{Conclusion}
\label{sec:conclusion}

We have shown that factual forecasting accuracy is not sufficient for building HVAC
control, and that foundation models must be structurally corrected to be
control-ready. ThermoForce achieves this by keeping a time-series foundation model
frozen as a passive free-response prior and learning a compact, monotone,
physics-structured forced-response operator for the causal effect of control. The
resulting interventional world model attains the lowest intervention-effect error
and correct effect sign where observational, covariate-TSFM, and distillation
baselines fail, preserves few-shot factual accuracy, and, embedded in MPC, reduces
BOPTEST thermal discomfort by 33--84\% while reducing energy, using a frozen backbone
and 195 trainable parameters. We believe the free/forced separation is a general
recipe for turning powerful passive predictors into control-ready world models for
energy systems.

\section*{Data availability}
The building datasets used are publicly available (REFIT, EnergyPlus reference
models, BOPTEST). Code and generated intervention data will be released upon
publication.

\bibliographystyle{elsarticle-num}
\bibliography{refs}

\end{document}

%% file: tables/tab_positioning.tex
\begin{tabular}{lcccccc}
\toprule
Approach & \rotatebox{90}{Uses TSFM} & \rotatebox{90}{TSFM frozen} & \rotatebox{90}{Explicit interv. op.} & \rotatebox{90}{Physics struct.} & \rotatebox{90}{Counterfactual eval.} & \rotatebox{90}{Closed-loop MPC} \\
\midrule
TSFM forecasting (Chronos, TimesFM) & \checkmark & \checkmark &  &  &  &  \\
TSFM for building energy (Park; Mulayim) & \checkmark &  &  &  &  &  \\
Physics-constrained thermal / RC &  &  &  & \checkmark &  & \checkmark \\
RC/ANN-MPC on BOPTEST (Zheng'24) &  &  &  & \checkmark &  & \checkmark \\
Differentiable predictive control (Drgo\v{n}a) &  &  &  & \checkmark &  & \checkmark \\
ThermoStill (e-Energy'26) & \checkmark &  &  & \checkmark &  & \checkmark \\
\textbf{ThermoForce (ours)} & \checkmark & \checkmark & \checkmark & \checkmark & \checkmark & \checkmark \\
\bottomrule
\end{tabular}

%% file: tables/tab_thermostill.tex
\begin{tabular}{lll}
\toprule
Dimension & ThermoStill (e-Energy'26) & \textbf{ThermoForce (ours)} \\
\midrule
TSFM role & Teacher / distillation source & Frozen passive prior \\
Model structure & Single distilled dynamics net & Free + forced superposition \\
HVAC control effect & Implicit inside one net & Explicit interventional operator \\
Physics structure & Learned surrogate & RC + sign/monotone forced resp. \\
Main purpose & Efficient thermal model for MPC & Counterfactual-valid world model \\
Core evidence & MPC thermal-dynamics utility & Factual + effect + closed-loop \\
Key claim & TSFM can help HVAC MPC & TSFM must be structurally corrected \\
\bottomrule
\end{tabular}

%% file: tables/tab_intervention_master.tex
\begin{tabular}{llccc}
\toprule
Mode & Method & Factual RMSE & \textbf{Effect RMSE} & Sign acc. \\
\midrule
\multirow{6}{*}{Heating (Chicago)} & TSFM raw & 0.817 & 1.439 & 0.00 \\
 & TSFM + covariates & 0.817 & 0.905 & 0.78 \\
 & Observational RC & 0.761 & 1.360 & 0.93 \\
 & Distillation surrogate & -- & 4.010 & 0.76 \\
 & ThermoForce (unconstr.\ gain) & 0.817 & 0.759 & 0.85 \\
 & \textbf{ThermoForce (monotone)} & 0.817 & 0.751 & 0.85 \\
\midrule
\multirow{6}{*}{Cooling (Phoenix)} & TSFM raw & 0.749 & 0.640 & 0.00 \\
 & TSFM + covariates & 0.749 & 0.331 & 0.68 \\
 & Observational RC & 1.391 & 0.640 & 0.01 \\
 & Distillation surrogate & -- & 0.552 & 0.90 \\
 & ThermoForce (unconstr.\ gain) & 0.749 & 0.331 & 0.96 \\
 & \textbf{ThermoForce (monotone)} & 0.749 & 0.245 & 0.95 \\
\bottomrule
\end{tabular}

%% file: tables/tab_main_refit.tex
\begin{tabular}{lcccc}
\toprule
& \multicolumn{2}{c}{6\,h horizon} & \multicolumn{2}{c}{24\,h horizon} \\
\cmidrule(lr){2-3}\cmidrule(lr){4-5}
Method & RMSE (K) & beats TSFM & RMSE (K) & beats TSFM \\
\midrule
Persistence & 0.992 & -- & 1.263 & -- \\
TSFM raw (zero-shot) & 0.458 & -- & 0.634 & -- \\
RC full (weather-oracle) & 0.663 & 6/19 & 0.868 & 5/19 \\
RC persist-weather & 0.703 & 4/19 & 1.241 & 3/19 \\
\textbf{ThermoForce (ours)} & 0.404 & 11/19 & 0.628 & 9/19 \\
\bottomrule
\end{tabular}

%% file: tables/tab_covariate.tex
\begin{tabular}{lcccc}
\toprule
& \multicolumn{2}{c}{6\,h} & \multicolumn{2}{c}{24\,h} \\
\cmidrule(lr){2-3}\cmidrule(lr){4-5}
Method & RMSE & FC wins & RMSE & FC wins \\
\midrule
TSFM raw (zero-shot) & 0.458 & 11/19 & 0.634 & 9/19 \\
TSFM + covariates (ridge) & 0.594 & 15/19 & 0.703 & 14/19 \\
TSFM + covariates + AR & 0.642 & 17/19 & 0.959 & 16/19 \\
\textbf{ThermoForce (ours)} & 0.404 & -- & 0.628 & -- \\
\bottomrule
\end{tabular}

%% file: tables/tab_closedloop.tex
\begin{tabular}{llccccc}
\toprule
Scenario & Controller & Energy & Discomfort & Cost & $\Delta$E & $\Delta$Dis. \\
\midrule
\multirow{3}{*}{Peak A (2 wk)} & Native baseline & 3.478 & 8.382 & 0.882 & -- & -- \\
 & RC-MPC & 3.128 & 5.208 & 0.793 & -10.1\% & -37.9\% \\
 & \textbf{ThermoForce-MPC} & 3.160 & 1.726 & 0.801 & -9.1\% & -79.4\% \\
\midrule
\multirow{3}{*}{Peak B (2 wk)} & Native baseline & 3.268 & 8.292 & 0.829 & -- & -- \\
 & RC-MPC & 2.994 & 0.832 & 0.759 & -8.4\% & -90.0\% \\
 & \textbf{ThermoForce-MPC} & 2.990 & 1.304 & 0.758 & -8.5\% & -84.3\% \\
\midrule
\multirow{3}{*}{Typical (2 wk)} & Native baseline & 1.773 & 9.446 & 0.449 & -- & -- \\
 & RC-MPC & 1.721 & 9.482 & 0.436 & -2.9\% & +0.4\% \\
 & \textbf{ThermoForce-MPC} & 1.707 & 6.352 & 0.433 & -3.7\% & -32.8\% \\
\bottomrule
\end{tabular}

%% file: tables/tab_ablation_operator.tex
\begin{tabular}{ccccc}
\toprule
& \multicolumn{2}{c}{Linear gain} & \multicolumn{2}{c}{Monotone-neural gain} \\
\cmidrule(lr){2-3}\cmidrule(lr){4-5}
Probe days & DR-down & DR-up & DR-down & DR-up \\
\midrule
1 & 0.445 & 0.570 & 0.448 & 0.486 \\
3 & 0.390 & 0.510 & 0.371 & 0.459 \\
7 & 0.386 & 0.497 & 0.313 & 0.453 \\
30 & 0.384 & 0.489 & 0.318 & 0.430 \\
\bottomrule
\end{tabular}

%% file: tables/tab_backbone.tex
\begin{tabular}{lcccc}
\toprule
Frozen backbone & Params & TSFM raw & ThermoForce & $\Delta$ (\%) \\
\midrule
chronos-bolt-tiny & 9\,M & 0.537 & 0.403 & -24.9 \\
chronos-bolt-mini & 21\,M & 0.523 & 0.409 & -21.8 \\
chronos-bolt-small & 48\,M & 0.510 & 0.396 & -22.4 \\
\bottomrule
\end{tabular}

%% file: tables/tab_efficiency.tex
\begin{tabular}{lcc}
\toprule
Component & Cost & Note \\
\midrule
Interventional operator & 195 params & trainable \\
Frozen TSFM backbone & 47.7\,M params & 0 trainable \\
RC identification (K=3\,d) & $\sim$24\,ms & per building \\
Neural operator training & $\sim$5.6\,s & 300 epochs, per building \\
Forced rollout (24-step) & $\sim$2.1\,ms & per forecast \\
TSFM free forecast (512$\to$24) & $\sim$30\,ms & per forecast \\
\bottomrule
\end{tabular}